\documentclass[runningheads]{llncs}
\usepackage{graphicx}

\usepackage[ruled,vlined]{algorithm2e}
\usepackage{algorithmic}

\usepackage{color,soul}

\begin{document}
\title{Graph analytics workflows enactment \\ on just in time data centres \\ Position Paper \thanks{This work is funded by the project GALILEAN, LIRIS intergroup collaboration  \url{https://galilean-project.github.io}.}}
\titlerunning{Analysing neuronal streams}
%
\author{
Ali Akoglu\inst{1} 
\and
José-Luis Zechinelli-Martini\inst{2} 
\and
Hamamache Kheddouci\inst{3} \and
Genoveva Vargas-Solar\inst{4} 
}
\authorrunning{A. Akoglu et al.}
%
\institute{
University of Arizona, USA \\
\email{akoglu@arizona.edu} 
\and
Fundación Universidad de las Américas Puebla, Mexico \\
\email{joseluis.zechinelli@udlap.mx} 
\and
Université Claude Bernard Lyon 1, LIRIS, France \\
\email{hamamache.kheddouci@univ-lyon1.fr} 
\and
French Council of Scientific Research, LIRIS, France \\
\email{genoveva.vargas-solar@liris.cnrs.fr} 
}
\maketitle              
\begin{abstract}

This paper discusses our vision about  multirole-capable decision-making systems across a broad range of Data Science (DS) workflows working on graphs through disaggregated data centres. Our vision is that an alternative is possible, to work on a disaggregated solution for the provision of computational services under the notion of a disaggregated data centre. 
We define this alternative as a virtual entity that dynamically provides resources crosscutting the layers of edge, fog and data centre according to the workloads submitted by the workflows and their Service Level Objectives. 

\end{abstract}
%
%
%
\section{Introduction}

Data collections can be structured as networks that have interconnection rules determined by the variables characterising each observation. The graph is a powerful mathematical concept with associated operations that can be implemented through efficient data structures and exploited by applying different algorithms. Note that relations among observations and interconnection rules are often not explicit, and it is the role of the analytics process to deduce, discover and eventually predict them. 

When the graphs become large and even too large
the algorithms used to process, explore and analyse them become costly in execution time, even if several cores are used. In this case, given the characteristics of the algorithms, communication is also likely to be costly. So workflows that exploit graphs become gluttonous consumers of computing resources. 
Our work comes into the scene in this context; we are interested in the execution conditions of graph processing workflows.

Data Science (DS) workflows pose unique challenges due to the growing complexity of processing and querying methods for big-data applications increasingly governed by analytics operations, machine learning-based workflows and models. 

Considering DS workflows' complexity, heterogeneity, and dynamic behaviour, it is impossible to produce a timely computing system solution in response to many dynamically arriving streaming applications and associated queries over complex graphs with potentially millions of nodes. Meeting performance requirements of large-scale DS workflows with tasks applying greedy operations applied on graphs across the entire dynamic system execution space is a daunting task without a clear understanding of the dependencies from available data sources to information extraction algorithms, from available information to decision algorithms, from algorithms to performance requirements, and from heterogeneous computing resources to performance capabilities.

Graph processing and analysis workflows consist of tasks that include:
\begin{itemize}
    \item  deploying or retrieving graphs which are often distributed over an execution environment, 
    \item  applying  algorithms of varying complexity in a distributed way, and
    \item   retrieving the results and making them available to other processes or to the end-users.
\end{itemize}
In terms of infrastructure, the execution takes place on often heterogeneous architectures that provide computing services with different capacities to execute them. 
From this point of view, it is possible to access computing solutions that range from  resource rich cloud based infrastructures all the way to the edge based power and resource limited resources.

In the current context,  workloads are typically greedily delegated to the cloud or data centres. Still, the computing resources residing on these architectures cannot be composed in an elastic and integrated way to build ad hoc execution environments on the fly.
Our work is in the context of approaches to designing alternative architectures to provide computing, storage, and memory resources that are more elastic and lightweight than greedy based approaches.

In addition to graph-based workflows, today's environments favour high-performance cloud-based platforms as a means to outsource their execution completely. These monolithically designed platforms provide various infrastructure services with heterogeneous computing resources with different capabilities to design, execute and maintain workflows.

Our vision is that it is possible to alternatively provide computational services under the notion of a disaggregated data centre. So a virtual entity that dynamically provides resources that touch the edge, fog and data centre according to the workloads submitted by the workflows and their Service Level Objectives (SLOs). 
Therefore, this paper discusses our vision about multirole-capable decision-making systems across a broad range of DS workflows working on graphs through an agile, autonomous, composable, and resilient "Just-in-Time Architecture" for DS Pipelines (JITA-4DS) \cite{akoglu2021putting}. 

Accordingly, the remainder of the paper is organised as follows.
Section \ref{sec:relatedwork} discusses related work regarding existing disaggregated data centres approaches and data science workflow execution. 
Section  \ref{sec:approach} describes our vision and research challenges about graph data science workflows and execution on disaggregated data centres. Section \ref{sec:jita4DS} describes the general lines of how to build just in time virtual data centres for executing data science workflows. Section \ref{sec:conclusion} concludes the paper and discusses  future work.


\section{Related Work} \label{sec:relatedwork}
In general, querying techniques can be categorised across two families: (i) the first concerns querying as we know it in databases and information retrieval; 
(ii) the second, a family where workflows, namely Data Science (DS)  workflows, explore and analyse the data to profile them quantitatively either with modelling, prediction, or recommendation purposes. The results of queries have an associated degree of error, and they may not only be data but also queries or data samples and models.

DS workflows need specialised architectures because of their size, dynamic behaviour, and nonlinear scaling and relatively unpredictable growth  with respect to their inputs being processed. Existing IT architectures are not designed to provide an agile infrastructure to keep up with the rapidly evolving next-generation mobile, big data, and data science workflows demands. They require continuous provisioning and re-provisioning of DC resources \cite{chen19acm,kannan19eurosys,xu2015enreal} given their dynamic and unpredictable changes in the SLOs (e.g., availability response time, reliability, energy).  

Existing DS environments are "one-fits-all" cloud systems that can manage and query data with different structures through built-in or user-defined operations integrated into imperative or SQL like solutions. They are provided by major vendors like Google, Amazon, IBM and Microsoft. They address the analytics and data management divide with integrated backends for efficient execution of analytics activities workflows, allocating the necessary infrastructure (CPU, FPGA, GPU, TPU) and platform (Spark, Tensorflow) services. These environments provide resources for executing DS tasks requiring storage and computing resources. DS workflows evolve from in-house executions into deployment phases on the cloud. Therefore, they need underlying elastic architectures that can provide resources at different scales. Disaggregated data centres solutions seem promising for them. Our work addresses the challenges implied when coupling disaggregated solutions with DS workflows.

\section {Graph DS workflows execution on disaggregated data centres}
\label{sec:approach}
 The research we propose aims to study the execution of DS workflows addressing graph analytics focusing on data processing, transmission and sharing across several resources. 
 We identify research challenges to study the execution of graph analysis workflows concerning the processing, transmission and sharing of data and different resources. Our hypothesis is that it is possible to schedule its tasks on a Virtual Data Science Centre (VDS) given a workflow.
We organise our study around three research questions:
\begin{itemize}
    \item [R1] Is it possible to adopt a database approach and draw on query evaluation to define the execution plan(s) of workflows taking into account the data distribution/execution load?
    
    \item [R2] How and according to which metrics  can we estimate the resource required by each task depending on the algorithm it calls and the volume of data to be processed?
    
    \item [R3] According to which strategies  can we estimate and configure the VDS according to a given workflow execution plan?

\end{itemize}
Given the difficulty of the problem, we propose to adopt  a three step data management and processing  methodology as summarised below.

\paragraph{Disaggregated Data Centre.}
We start from the abstract idea of a disaggregated data centre as a possible configuration in the form of a virtual machine that provides computing, storage and RAM resources available on a Data Centre Building Block Pool. The needs of a workflow guide the configuration of VDS in terms of execution, monitoring and maintenance throughout its lifecycle.

\paragraph{Executing Data Science Workflows.}
The execution of DS workflows on graphs consists of data processing tasks to be scheduled on a disaggregated VDS. Our approach is to define execution, configuration and deployment plans that can guide the execution, to represent the strategies to allocate resources and calibrate a VDS  according to the characteristics of a given data science workflow. Therefore a first challenge to address is to rewrite DS workflows into these plans. The objective is to define data dependencies among tasks and the control flow to adopt for executing them considering the distribution of the data/execution workloads.

Our study has started from pipelines using analytics graph algorithms to answer community detection problems like page rank, Louvain, more mathematical models applied to matrices representing graphs (run in the LNS, Mexico) according to previous work \cite{bouhenni2021survey,brighen2020distributed}. Our focus will be on the characterization of DS workflows considering (i) the type of graph processing algorithms they address; (ii) the characteristics of the graphs (data) processed and results through these algorithms.

Designing and rewriting strategies for generating execution plans implies the definition of metrics to estimate costs and SLOs in the different phases of the workflow execution cycle. The execution must be guided by dynamic and elastic provisioning of resources. The challenge to address is to estimate the resources requirements associated with each task of the execution plan according to the algorithm it calls and the data injection function estimating the volume of data to process. In this context, experiments are essential to guide and validate the proposals. 

We have focused on defining the right metrics for estimating the requirements of target DS workflows as presented in our previous work~\cite{akoglu2021putting}. We describe the SLO objectives of given DS workflows on graphs that should be fulfilled at execution time. 

\paragraph{Estimating and configuring  initial VDS workflows.}
Our focus is in proposing a DS workflow rewriting strategy that will generate an ad hoc execution specification including (i) tasks to be executed by the workflow (classic execution plan); (ii) the corresponding specification of the underlying VDS workflow architecture (configuration plan) and (iii) the deployment plan defining the distribution of the processes from the edge to the VDS workflow. DS workflows introduce other challenges like weaving data preparation, fragmentation, and analytics operations where data dependencies and requirements across tasks must be fine-tuned and modelled.


\section {Towards Just in Time Virtual Data Centres for Data Science Workflows}
\label{sec:jita4DS}
Our research investigates architectural support, system performance metrics, resource management algorithms, and modelling techniques to enable the design of composable (disaggregated) DCs. 
The goal is to design an innovative composable “Just in Time Architecture” for configuring DCs for Data Science Pipelines (JITA-4DS) and associated resource management techniques \cite{akoglu2021putting}. DCs utilize a set of flexible building blocks that can be dynamically and automatically assembled and re-assembled to meet the dynamic changes in workload’s SLOs of current and future DC applications.
DCs under our approach are composable based on vertical integration of the application, middleware/operating system, and hardware layers 
customized dynamically to meet application SLO (application-aware management). 
Thus,  DCs configured using JITA-4DS provide ad-hoc environments efficiently and effectively meeting the continuous evolution of requirements of data-driven applications or workloads (e.g., data science pipelines). 

A DC is based on a novel application-aware VDC Management  system by dynamically invoking the appropriate actions to change the current VDC configuration to meet its objectives at runtime. 
To assess disaggregated DC's, we study how to model and validate their performance in large-scale settings. We rely on our novel model-driven resource management heuristics ~\cite{kumbhare2017value,kumbhare2020dynamic,kumbhare2020value} based on metrics that measure a service's value for achieving a balance between competing goals (e.g., completion time and energy consumption). Our focus is on defining new system performance measures that combine objectives, such as execution time and energy use, that dynamically change during the day.

Initially, we propose a hierarchical modelling approach that integrates simulation tools and models. 
 Results can be used for developing benchmarks that accurately characterize the requirements and SLOs of next-generation DC applications.

\begin{figure}[ht!] \centering
\includegraphics[width=0.95\textwidth]{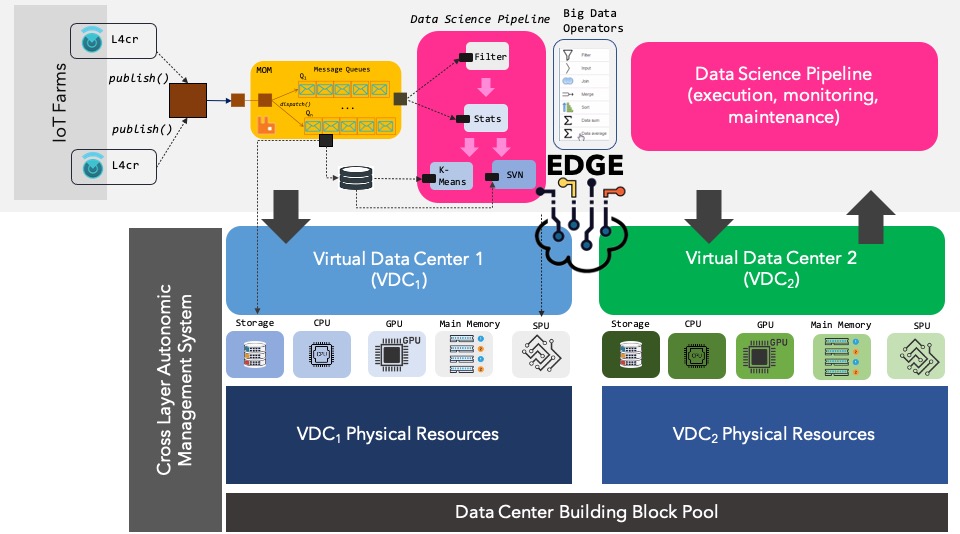}
\caption{Just in Time Architecture for Data Science Pipelines - JITA-4DS.}\label{fig:jita4ds}
\end{figure}

The Just in Time Architecture for Data Science Pipelines (JITA-4DS), illustrated in Figure \ref{fig:jita4ds}, is a cross-layer management system that is aware of both the application characteristics and the underlying infrastructures to break the barriers between applications, middleware/operating system, and hardware layers. Vertical integration of these layers is needed to build a customizable VDC to meet the dynamically changing data science pipelines' performance, availability, and energy consumption requirements. 

JITA-4DS  can build a VDC that can meet the application SLO for execution performance and energy consumption to execute data science pipelines. Then, the selected VDC is mapped to a set of heterogeneous computing nodes such as GPPs, GPUs, TPUs, special-purpose units (SPUs) such as ASICs and FPGAs, along with memory and storage units. 

\section{Conclusions and Future Work}\label{sec:conclusion}
This paper introduced our vision and research position regarding the design of just in time architectures for providing disaggregated resources for the execution of graph analytics workflows.  The originality of our research program is promoting the
provision of resources holistic system through intelligent resource management. This holistic system integrates three elements, graph processing models, associated computational resources, autonomous execution of complex and dynamic workflows.

From a more general point of view, three aspects characterise the approach.
 Its pioneering and promising aspect tackles the design of disaggregated data centres to address execution environments' design for data science workflows applied to graphs.
 
 We have described the general characteristics of our current results regarding  JITA-4DS. This virtualised architecture provides a disaggregated data centre solution ad hoc for executing DS workflows requiring elastic access to resources. DS workflows process graphs coordinating operators implemented by services deployed on edge. Since operators can implement greedy tasks with computing and storage requirements beyond those residing on edge, they interact with VDC services. We have set the first simulation setting to study resources delivery in JITA-4DS.

We are currently addressing challenges of VDCs management on simpler environments, on cloud resource management heuristics, (e.g., \cite{kumbhare2017value,machovec2017utility,kumbhare2020dynamic,kumbhare2020value}), 
big data analysis, 
and data mining for performance prediction.
To simulate, evaluate, analyze, and compare different heuristics, we will build 
simulators for simpler environments and combine open-source simulators for different levels of the JITA-4DS hierarchy.

Disaggregated approaches for providing data centres resources are emerging as a promising topic discussed in panels at major conferences and by leading scientists and companies. For the time being, approaches address the communication layers, but the wave is starting to touch computing and storage and platform levels. We have a first proposal for including the edge because of the characteristics of DS workflows.

To conclude, we believe that reasoning about the design and provision of alternatives to data science execution environments under a disaggregated perspective is pioneering and promising. Supporting this kind of exploratory project can encourage  digital independence  on the  way data science experimentation is enacted and can provide solutions beyond the lab walls. 


\bibliographystyle{splncs04}
\bibliography{biblio}
\end{document}